\documentclass[prc,twocolumn,superscriptaddress,nofootinbib]{revtex4-1}
\usepackage{newtxtext}
\usepackage[varvw,bigdelims]{newtxmath}
\usepackage[colorlinks=true,allcolors=blue]{hyperref}
\usepackage{graphicx}
\usepackage{bm}
\usepackage{amsmath}
\usepackage{xcolor}

\begin{document}
\title{\boldmath The $\eta^\prime N$ interaction from the $\eta^\prime p$ correlation function} 

\author{Natsumi Ikeno}
\email{ikeno@maritime.kobe-u.ac.jp}
\affiliation{Graduate School of Maritime Sciences, Kobe University, Kobe 658-0022, Japan}

\begin{abstract}
We evaluate for the first time the $\eta^\prime p$ femtoscopic correlation function to study the $\eta^\prime N$ interaction. 
We find it extremely sensitive to the value of the $\eta^\prime p$ scattering length, for which at present there exists only very limited information, not even knowing its sign. 
The measurement of this correlation function would provide much valuable information on the $\eta^\prime N$ interaction, which could then also be used to settle the issue of possible $\eta^\prime$ nucleus bound states, an issue attracting much attention in the nuclear physics community.

\end{abstract}

\date{\today}

\maketitle
\section{Introduction}
The meson–nucleus bound systems, such as mesic atoms and mesic nuclei, are very intriguing objects to study aspects of the strong interaction~\cite{Batty:1997zp,Friedman:2007zza,Hayano:2008vn,Metag:2017yuh}. 
For instance, the study of deeply bound pionic atoms, the pion–nucleus bound system, has provided valuable information on the in-medium pion properties and the restoration of chiral symmetry at finite density~\cite{Kolomeitsev:2002gc,Suzuki:2002ae,                Jido:2008bk,Ikeno:2011mv,Ikeno:2011aa,Yamazaki:2012zza,Ikeno:2013wza,Goda:2013npa,Ikeno:2015ioa,piAF:2017nha,Friedman:2019zhc,Hirenzaki:2022dpt,Itahashi:2023boi,Ikeno:2022foa,PiAF:2022gvw}. 
Various meson–nucleus systems have also been studied intensively both theoretically and experimentally, e.g.~\cite{Ikeno:2011rf,Xie:2016zhs,Ikeno:2017xyb,Skurzok:2018paa,J-PARCE15:2020gbh,J-PARCE62:2022qnt,Tani:2020yvr,Cobos-Martinez:2022fmt,Yamagata-Sekihara:2024zrk}.

In particular, the study of an $\eta^\prime(958)$ ($\eta^\prime$) meson--nucleus bound state, $\eta^\prime$ mesic nucleus, has attracted considerable attention because the $\eta^\prime$ meson is significantly heavier than the other light pseudoscalar mesons. 
The exceptionally large mass of the $\eta^\prime$ meson is theoretically understood by the interplay between chiral symmetry breaking and the $U_A(1)$ anomaly~\cite{Jido:2011pq,Nagahiro:2012aq} (see also recent reviews~\cite{Bass:2018xmz,Bass:2021rch} on the $\eta$ and $\eta^\prime$ mesons). Therefore, the $\eta^\prime$ mesic nucleus is expected to be a unique system to probe the effects of the $U_A(1)$ anomaly at finite density. 

To search for the $\eta^\prime$ mesic nucleus, 
the structure and formation of the $\eta^\prime$ mesic nucleus have been studied theoretically in Refs.~\cite{Nagahiro:2004qz,Nagahiro:2006dr,Nagahiro:2011fi,Jido:2011pq,Nagahiro:2012aq,Itahashi:2012ut,Miyatani:2016xfq,Jido:2018aew,Cobos-Martinez:2023hbp}. However, the existence has not been confirmed experimentally yet~\cite{n-PRiMESuper-FRS:2016vbn,e-PRiMESuper-FRS:2017bzq,LEPS2BGOegg:2020cth}.
In Refs.~\cite{n-PRiMESuper-FRS:2016vbn,e-PRiMESuper-FRS:2017bzq}, the formation of the $\eta^\prime$ mesic nucleus by the $^{12}$C($p,d$) reaction was attempted experimentally, and it was found that clear signals of the $\eta^\prime$--nucleus bound state were not observed due to large background contributions.
In order to reduce the background contributions, the semi-exclusive $^{12}$C($p,dp$) reaction has recently been discussed in Refs.~\cite{Fujioka:2014nfa,Super-FRS:2015sjd,Higashi2015,ikeno:2024slo}, and the experimental analysis is currently in progress~\cite{itahashi_private}.
As for related topics, some experimental studies have also been performed to obtain the in-medium $\eta^\prime$ properties and the $\eta^\prime$--nucleus interaction in Refs.~\cite{CBELSATAPS:2012few,CBELSATAPS:2016qdi,Friedrich:2016cms,CBELSATAPS:2018sck}. 
We are increasingly obtaining constraints on the $\eta^\prime$--nucleus interaction.

The knowledge of the $\eta^\prime N$ interaction is essential as a fundamental piece for the study of the $\eta^\prime$ mesic nucleus.
To understand the $\eta^\prime N$ interaction as well as the in-medium $\eta^\prime$ properties, many studies have been carried out with various theoretical models such as the chiral coupled channels approach~\cite{Oset:2010ub,Nagahiro:2011fi,Bruns:2019fwi}, the linear sigma model~\cite{Sakai:2013nba,Sakai:2016vcl}, the quark-meson coupling model~\cite{Cobos-Martinez:2023hbp}, and so on.
Although experimental determinations of the scattering length for the $\eta^\prime N$ interaction have been reported~\cite{Anisovich:2018yoo,Czerwinski:2014yot}, it has still a large ambiguity, and even the sign of the $\eta^\prime N$ scattering length has not yet been clearly determined from the experimental analysis.

Recently, the femtoscopic technique has been a very powerful tool for probing the two-particle interaction through the measurement of the two-particle correlation function~\cite{STAR:2014dcy,ALICE:2017jto,STAR:2018uho,ALICE:2018ysd,ALICE:2019hdt,ALICE:2019eol,ALICE:2019buq,ALICE:2019gcn,ALICE:2020mfd,ALICE:2021szj,ALICE:2021cpv,Fabbietti:2020bfg,ALICE:2022enj}. 
Numerous studies have been carried out to consider the femtoscopic correlation function between various particle pairs~\cite{Morita:2014kza,Ohnishi:2016elb,Morita:2016auo,Hatsuda:2017uxk,Mihaylov:2018rva,Haidenbauer:2018jvl,Morita:2019rph,Kamiya:2019uiw,Kamiya:2021hdb,Kamiya:2022thy,Liu:2023uly,Vidana:2023olz,Albaladejo:2023pzq,Ikeno:2023ojl,Molina:2023jov,Feijoo:2024bvn,Jinno:2024tjh,Kamiya:2024diw,Encarnacion:2024jge,Liu:2023wfo,Liu:2024uxn,Liu:2024nac,Encarnacion:2025lyf,Ikeno:2025bsx}. 
For instance, in Ref.~\cite{Molina:2023jov}, the $\eta N$ correlation function was studied to understand the property of the $N^*(1535)$ state, which is related to the $\eta N$ interaction and is dynamically generated with the coupled channels approach including the $\eta N$ channel.

In this article, we focus on the $\eta^\prime N$ femtoscopic correlation function to study the scattering length for $\eta^\prime N$ interaction.
In the femtoscopic study, the behavior of the correlation function is used to obtain the two-particle interaction~\cite{Fabbietti:2020bfg,Liu:2023uly}. 
%
%
We expect that the $\eta^\prime N$ femtoscopic correlation function can provide valuable new information on the $\eta^\prime p$ scattering length, including its sign, which has not yet been clearly determined by other experiments.

This paper is organized as follows. In Sec.~\ref{sec:formalism}, we explain the formalism to obtain the $\eta^\prime N$ interaction with the coupled channel approach and the $\eta^\prime p$ correlation function. In Sec.~\ref{sec:result}, we show our numerical results, and Sec.~\ref{sec:discussion} is devoted to summary and conclusion.

\section{Formalism}\label{sec:formalism}

\subsection{The $\eta^\prime N$ interaction}
We use the coupled-channel approach based on Ref.~\cite{Oset:2010ub} to obtain the $\eta^\prime N$ interaction. In this article, we consider the $K \Lambda$, $K \Sigma$, $\eta N$, and $\eta^\prime N$ channels. The $\pi N$ channel is not included, as its energy is far below the $\eta^\prime N$ threshold, and its contribution is negligible as in Ref.~\cite{Molina:2023jov}.

For simplicity, we follow the same formula as explained in Ref.~\cite{Molina:2023jov} for the $N^*(1535)$ state which is dynamically generated by the interaction of coupled channels. 
We consider the same coupled channels as in Ref.~\cite{Molina:2023jov,Inoue:2001ip} and additionally include the $\eta^\prime N$ channel based on Ref.~\cite{Oset:2010ub}.
Then, we take the coupled channels considered as, 
\begin{equation}
 K^0 \Sigma^+, \ K^+ \Sigma^0,  \ K^+ \Lambda, \ \eta p, \ \eta^\prime p.
\nonumber
\end{equation}
The interaction between these channels is given by
\begin{equation}
 V_{ij}=-\dfrac{1}{4f^2} C_{ij} (k^0+k^{\prime \,0});~~~~ f=93\, {\rm MeV},
\label{eq:Vij}
\end{equation}
with $k^0, k^{\prime \,0}$ the relativistic energies of the initial and final mesons in the center of mass system,
\begin{equation}
k^0 =\dfrac{s+m_1^2-M_1^2}{2\sqrt{s}};~~~~ k^{\prime \,0} =\dfrac{s+m_2^2-M_2^2}{2\sqrt{s}},
\label{eq:k0}
\end{equation}
where $m_1$ and $M_1$ are the masses of the initial meson and baryon, and $m_2$ and $M_2$ are the same for the final ones.
The physical meson fields of $\eta$ and $\eta^\prime$ are related to the SU(3) singlet ($\eta_1$) and octet ($\eta_8$) fields by the relation,
\begin{equation}
\begin{aligned}
 \eta & = \cos\theta_p \eta_8 - \sin\theta_p \eta_1  ,\\
 \eta^\prime & = \sin\theta_p \eta_8 + \cos \theta_p \eta_1,
 \end{aligned}
\label{eq:mixing}
\end{equation}
with the $\eta_1$--$\eta_8$ mixing angle $\theta_p$. We take the value of $\theta_p = -14.34^\circ$~\cite{Ambrosino:2009sc,Oset:2010ub} in this article.
Considering the mixing of Eq.~\eqref{eq:mixing}, the coefficients $C_{ij}$ in Eq.~\eqref{eq:Vij} are given in Table \ref{tab:Cij}.

\begin{table}[!tb]
\centering
\caption{The coefficients $C_{ij}$ of Eq.~\eqref{eq:Vij}.}
\begin{tabular}{c| c c c c c c c }
\hline 
$C_{ij}$         &~~ $K^0 \Sigma^+$ ~~ & ~~ $K^+ \Sigma^0$~~  & ~~ $K^+ \Lambda$ ~~  &  $\eta p$  &  $\eta^\prime p$ \\
\hline
$K^0 \Sigma^+$   & $1$&  $\sqrt{2}$      &  $0$    &  $-\sqrt{\frac{3}{2}} \cos\theta_p$ &  $-\sqrt{\frac{3}{2}} \sin\theta_p$ \\
$K^+ \Sigma^0$   &    & $0$   &  $0$ &   $-\frac{\sqrt{3}}{2} \cos\theta_p$ &   $-\frac{\sqrt{3}}{2} \sin\theta_p$ \\
$K^+ \Lambda$    &    &  &  $0$    &  $-\frac{3}{2} \cos\theta_p$ &  $-\frac{3}{2} \sin\theta_p$ \\
$\eta p$         &    &   &     & $0$  & $0$ \\
$\eta^\prime p$         &    &   &     & $0$ & $0$ \\
\hline
\end{tabular}
\label{tab:Cij}
\end{table}

We take into account additional potential terms associated with the Lagrangian that couples the singlet meson to the baryons~\cite{Borasoy:1999nd}, as introduced in Ref.~\cite{Oset:2010ub}.
The additional potential terms between $\eta N$ and $\eta^\prime N$ are given by
\begin{equation} 
 \begin{aligned}
  &V_{\eta N, \eta N} = C \, \sin^2 \theta_p, \\
  &V_{\eta N, \eta^\prime N} = -C \, \sin\theta_p \cos\theta_p, \\
  &V_{\eta^\prime N, \eta^\prime N} = C \, \cos^2 \theta_p,
 \end{aligned}
\label{eq:Vadd}
\end{equation}
with 
\begin{equation}
 C = \frac{\alpha}{4 f^2} 2 m_{\eta^\prime} \, \frac{k^0 + k^{\prime \,0} }{2 M_N},
\label{eq:Pot_alpha}
\end{equation}
where $\alpha$ is an unknown parameter which determines the strength of the potentials in Eq.~\eqref{eq:Vadd} due to the coupling between $\eta_1$ and $N$. 
We consider the various values of $\alpha$ in this article to study the $\eta^\prime p$ scattering length.
We adopt the theoretical model described above since our primary goal is to test the sensitivity of the correlation function for $\eta^\prime p$ to its interaction.


{We note that in the present calculation we only adopt the $s$-wave interaction kernel of Ref.~\cite{Oset:2010ub} for the meson--baryon interaction in the strangeness $S=0$ sector for our purpose to test the sensitivity of the correlation function to the
$\eta^\prime p$ interaction, and, therefore, we do not include the contribution of resonances around $2$~GeV that in Ref.~\cite{Oset:2010ub} were incorporated through box diagrams coupling to vector--baryon (VB) states dynamically generated in their scheme. These mechanisms are not dominant in the immediate threshold region, yet they are not negligible. 
Moreover, higher-order terms of the chiral Lagrangian and higher partial waves, such as the $p$-wave components discussed in Refs.~\cite{CaroRamon:1999jf,Schopper:1988vrx,Inoue:2001ip,Bruns:2019fwi}, are absent in the present model. 
In particular, the authors of Ref.~\cite{CaroRamon:1999jf} incorporate $s$- and $p$-wave projections of a kernel derived from the SU(3) chiral Lagrangian up to next-to-leading order (NLO), and demonstrate that the $p$-wave contributions are essential to reproduce the experimental cross sections of the reactions $\pi^- p \to K^+ \Sigma^-$, $K^0 \Sigma^0$, $\eta n$, and $\pi^+ p \to K^+ \Sigma^+$ at energies near the $\eta'N$ threshold. 
The necessity of higher-order chiral corrections and higher partial waves in describing the $\pi N \to \eta N$ reaction at such energies was pointed out in Ref.~\cite{Inoue:2001ip}. 
In this context, $u$- and $s$-channel diagrams with a sizeable $p$-wave projection may provide additional contributions to the $\eta'N$ interaction. In Ref.~\cite{Bruns:2019fwi}, their $s$-wave components together with tree-level NLO terms were already taken into account. 
All these missing ingredients can modify the strength of the amplitudes involved in the calculation and introduce additional interplay between channels to some extent. 
In our study, the strength of the interaction is determined by the value of the potential parameter $\alpha$ in Eq.~\eqref{eq:Pot_alpha}, and thus these uncertainties may be partially absorbed into the choice of this parameter.
}

To obtain the scattering matrix $T$, we solve the Bethe-Salpeter (BS) equation in coupled channels written in the schematic form as,
\begin{equation}
  T=[1-VG]^{-1} \, V,
\label{eq:BS}
\end{equation}
with $V$ the potentials of Eqs.~\eqref{eq:Vij} and~\eqref{eq:Vadd}, and $G$ the elements of the diagonal matrix of the loop function ${\rm diag}(G_i)$.
We regularize the loop function $G_i$ for each channel $i$ with a cutoff parameter $q_{\rm max}$ as,
\begin{align}
  G_i(s) = 2M_i & \int_{|{\bm q}| < q_{\rm max}} \frac{{\rm d}^3 q}{(2\pi)^3} \; \frac{\omega_i({\bm q}) + E_i({\bm q})}{2\,\omega_i({\bm q})\, E_i({\bm q})} \nonumber\\
  &\quad \cdot \frac{1}{s - [\omega_i({\bm q}) + E_i({\bm q})]^2 + i \varepsilon},
  \label{eq:G}
\end{align}
with $\omega_i({\bm q})=\sqrt{m_i^2 + {\bm q}^2 }$, $E_i({\bm q})=\sqrt{ M_i^2 + {\bm q}^2 }$, where $m_i$ and $M_i$ are the masses of the meson and baryon in the loop.
We basically use a value of $q_{\rm max}= 630$~MeV, as used in the study of $N^*(1535)$~\cite{Molina:2023jov}.

We follow the treatment of $G_i$ described in Ref.~\cite{Song:2024yli} and we only take into account the real part of $G_i$ with negative values, setting it to zero if it becomes positive.
We consider the exact value of $\text{Im}\,G_i$ as
\begin{equation}
\text{Im}\, G_i = - 2 M_i \frac{1}{8 \pi \sqrt{s}} q_{i \text{on}} 
\label{eq:ImG}
\end{equation}
with
\begin{equation}
q_{i \text{on}}=\frac{\lambda^{1/2}\left(s, m_i^2, M_i^2\right)}{2 \sqrt{s}},
\label{eq:q_on}
\end{equation}
where $\lambda$ is the K\"{a}llen function $\lambda(a,b,c) = a^2 + b^2 + c^2 - 2ab -2bc -2ca$.

The $\eta^\prime p$ scattering length $a_{\eta^\prime p}$ is calculated from the scattering amplitude $T_{\eta^\prime p \to \eta^\prime p}$ at the $\eta^\prime p$ threshold as~\cite{Oset:2010ub,Sakai:2022xao}\footnote{We follow the prescription $\displaystyle (f^{\text{Q}}(k))^{-1} \simeq \frac{1}{a}+\frac{1}{2} r_0 k^2 -ik$  where $f^{\text{Q}}(k)$ is the ordinary Quantum Mechanical scattering amplitude in textbooks and is related to $T$ defined in Eq.~\eqref{eq:BS} as $\displaystyle T \equiv - \frac{8 \pi \sqrt{s}}{2M} f^{\text{Q}}(k) $. }
\begin{equation}
 a_{\eta^\prime p} = -\frac{1}{4\pi} \frac{M_p}{m_{\eta^\prime} + M_p} ~ T_{\eta^\prime p \to \eta^\prime p},
\end{equation}
where the positive (negative) sign of the $a_{\eta^\prime p}$ corresponds to the attractive (repulsive) interaction.

\subsection{ The $\eta^\prime p$  correlation function}

We follow the formalism in Ref.~\cite{Vidana:2023olz}, which is a slight modification of the formalism based on the Koonin-Pratt method~\cite{Koonin:1977fh}, to take into account the finite range of the interaction to be consistent with the use of the cut-off regularization of the loop functions.
The $\eta^\prime p$ correlation function $C_{\eta^\prime p}$ for the coupled channel calculation is given by
\begin{align}
  C_{\eta^\prime p} (p) = 1 
  & + 4\pi \theta(q_{\rm max}-p) \int_0^{\infty} dr \, r^2 S_{12}(r)  \nonumber\\
  & \cdot \Bigg\{ \Big|j_0(p r) 
  + T_{\eta^\prime p, \eta^\prime p}(\sqrt{s})\; \tilde{G}^{(\eta^\prime p)}(r; s)\Big|^2 \nonumber\\
  &\quad + \Big|T_{K^0 \Sigma^+, \eta^\prime p}(\sqrt{s})\; \tilde{G}^{(K^0\Sigma^+)}(r; s)\Big|^2 \nonumber\\
  &\quad + \Big|T_{K^+ \Sigma^0, \eta^\prime p}(\sqrt{s})\; \tilde{G}^{(K^+\Sigma^0)}(r; s)\Big|^2 \nonumber\\
  &\quad + \Big|T_{K^+ \Lambda, \eta^\prime p}(\sqrt{s})\; \tilde{G}^{(K^+ \Lambda)}(r; s)\Big|^2 \nonumber\\
  &\quad + \Big|T_{\eta p, \eta^\prime p}(\sqrt{s})\; \tilde{G}^{(\eta p)}(r; s)\Big|^2 
  - j_0^2 (p r)\Bigg\},  
  \label{eq:CF}
\end{align}
where $T_{ij}$ with the indices $i$ and $j$ representing the meson-baryon channles are the elements of the scattering matrix of the coupled channels obtained from Eq.~\eqref{eq:BS}, $j_0(pr)$ is the spherical Bessel function, 
and $p$ is the momentum of the particles in the rest frame of the $\eta^\prime p$ pair,
\begin{equation}
\label{eq:p_etap}
  p =\frac{\lambda^{1/2}(s, m_{\eta^\prime}^2, M_p^2)}{2 \sqrt{s}}.
\end{equation} 
In Eq.~\eqref{eq:CF}, each inelastic channel has 
{a production weight ($\omega^{\text{prod}}_{i}$,~$i = K^0 \Sigma^+, K^+ \Sigma^0, K^+ \Lambda, \eta p$)} which in principle has to be obtained from the experiment. We have put the weight to be 1 in Eq.~\eqref{eq:CF}, as in the most theoretical papers.
We also calculate the $\eta^\prime p$ correlation function for the elastic channel only $C_{\eta^\prime p}^{\text{el}}$ as
\begin{align}
  C_{\eta^\prime p}^{\text{el}} (p) &= 1 
   + 4\pi \theta(q_{\rm max}-p) \int_0^{\infty} dr \, r^2 S_{12}(r)  \nonumber\\
  & \cdot \Bigg\{ \Big|j_0(p r) 
  + T_{\eta^\prime p, \eta^\prime p}(\sqrt{s})\; \tilde{G}^{(\eta^\prime p)}(r; s)\Big|^2   - j_0^2 (p r)\Bigg\}.
  \label{eq:CF2}
\end{align}
The function $S_{12}(r)$ appearing in Eqs.~\eqref{eq:CF} and \eqref{eq:CF2} is the source function, which takes into account the probability distribution of meson and baryon production as a function of relative distance $r$.
It is usually assumed to be a normalized Gaussian function as,
\begin{equation}
  S_{12}(r)= \dfrac{1}{(\sqrt{4\pi} R)^3}  \; \exp\left(-\frac{r^2}{4R^2}\right),
\label{eq:S12}
\end{equation}
with the source size $R$.
Typical values of $R$ are $1$~fm for $pp$ collisions and $5$~fm for heavy-ion $AA$ collisions.
In this article, we use $R=1$~fm for the calculations.

The function $\tilde{G}^{(i)}(r; s)$ in Eqs.~\eqref{eq:CF} and \eqref{eq:CF2} is defined as
\begin{align}
  \tilde{G}^{(i)}(r; s) = 2M_i  & \int \frac{{\rm d}^3 q}{(2\pi)^3} \, \frac{\omega_i({\bm q}) + E_i({\bm q})}{2\,\omega_i({\bm q})\, E_i({\bm q})} \nonumber\\
  &\quad \cdot \frac{j_0(q r)}{s - [\omega_i({\bm q}) + E_i({\bm q})]^2 + i \varepsilon},
  \label{eq:Gtild}
\end{align}
where $\omega_i({\bm q})$ and $E_i({\bm q})$ are the relativistic enegies defined in Eq.~\eqref{eq:G}.
$\tilde{G}^{(i)}(r; s)$ can be rewritten in a different form as~\cite{Albaladejo:2023wmv,Molina:2023jov}, 
\begin{align}
  \tilde{G}^{(i)}(r; s) = 2M_i  & \int \frac{{\rm d}^3 q}{(2\pi)^3} \, \frac{\omega_i({\bm q}) + E_i({\bm q})}{2\,\omega_i({\bm q})\, E_i({\bm q})} \nonumber\\
  &\quad \cdot \frac{j_0(q r) - j_0(q_{i \text{on}} r)}{s - [\omega_i({\bm q}) + E_i({\bm q})]^2 + i \varepsilon} \nonumber\\
+ \,2M_i \, j_0(q_{i \text{on}} r) & \int \frac{{\rm d}^3 q}{(2\pi)^3} \, \frac{\omega_i({\bm q}) + E_i({\bm q})}{2\,\omega_i(q)\, E_i(q)} \nonumber\\
  &\quad \cdot \frac{ 1 }{s - [\omega_i({\bm q}) + E_i({\bm q})]^2 + i \varepsilon}, 
  \label{eq:Gtild2}
\end{align}
where $q_{i \rm on}$ is the on-shell value of the momentum of each channel $i$, as given in Eq.~\eqref{eq:q_on}.
Since both the numerator and denominator approach zero as $q \to q_{\rm on}$,
the first term has no singularity and the limit is finite.
The second term is identical to $G_i(s)$ of Eq.~\eqref{eq:G} except for $j_0(q_{i \text{on}} r)$, and then we can perform the calculation in the same way as explained below Eq.~\eqref{eq:G}.

\section{Numerical Results}\label{sec:result}
First, we show the calculated $\eta^\prime p$ scattering lengths in Table~\ref{tab:scat} for the different values of the potential parameter $\alpha$ in Eq.~\eqref{eq:Pot_alpha}.  
We take various values of the potential parameter $\alpha$ to calculate the different scattering lengths following a similar procedure as in Refs.~\cite{Oset:2010ub,Nagahiro:2011fi}.
{
We note that, except for the $\eta N $ and $\eta^\prime N$ channels, the scattering lengths of the other channels change by less than about 5\% for the different $\alpha$ in Table~\ref{tab:scat}, since the parameter $\alpha$ depends on the potential strength of $\eta N $ and $\eta^\prime N$ channels.
}

So far, some experimental results on the $\eta^\prime p$ scattering length have been obtained as compiled in Refs.~\cite{Bass:2018xmz,Bass:2021rch}. 
The latest experimental result is shown in Ref.~\cite{Anisovich:2018yoo} as,
\begin{equation}
|a_{\eta^\prime p}| = (0.403 \pm 0.020 \pm 0.060)~\rm{fm}, 
\label{eq:Exp1}
\end{equation}
which is determined by the analysis of the $\gamma p \to \eta^\prime p$ reaction.
The experimental result by the $pp \to pp \eta^\prime$ reaction by COSY-11~\cite{Czerwinski:2014yot} is reported to be,
\begin{equation}
a_{\eta^\prime p} = 0^{+0.43}_{-0.43} + i0.37^{+0.40}_{-0.16}~\rm{fm},
\label{eq:Exp2}
\end{equation}
The two experimental results of Eqs.~\eqref{eq:Exp1} and \eqref{eq:Exp2} are consistent within the errors, however, the sign of the real part of the $\eta^\prime p$ scattering length is not clearly determined yet.

In Table~\ref{tab:scat}, we show the calculated results of the scattering length $a_{\eta^\prime p}$ by our model explained in the previous section for the various values of the parameter $\alpha$ defined in Eq.~\eqref{eq:Pot_alpha}. 
We find that the calculated absolute values of the scattering length $|a_{\eta^\prime p}|$ are consistent with the data in Eq.~\eqref{eq:Exp1} at $\alpha \sim 4.00$ and $\alpha \sim -0.50$. As for the real part of the observed scattering length in Eq.~\eqref{eq:Exp2}, almost all the calculated results in Table~\ref{tab:scat} are consistent with the value in Eq.~\eqref{eq:Exp2} within the errors.
On the other hand, the imaginary part of the observed scattering length in Eq.~\eqref{eq:Exp2} cannot be reproduced by the present model, as can be seen in Table~\ref{tab:scat}. The calculated result with $\alpha = -0.55$ provides the closest value to the observed $\text{Im}\, a_{\eta^\prime p}$ in the cases considered here.

There are also other theoretical results reported before.
The repulsive potential with $\alpha = 4.00$ in Table~\ref{tab:scat}  has a similar strength to the interaction constructed with the coupled channel approach in Ref.~\cite{Bruns:2019fwi}, where the $\eta^\prime p$ scattering length is obtained as $a_{\eta^\prime p} = -0.41 + i0.04$~fm. 
In the linear sigma model of Refs.~\cite{Sakai:2013nba,Sakai:2016vcl,Sakai:2022xao}, the $\eta^\prime p$ scattering length is obtained as $a_{\eta^\prime p} = +0.85$~fm, leading to a mass reduction of 80~MeV at the normal nuclear density. 
The $\eta^\prime p$ interaction is more attractive than that of our model with $\alpha = -0.55$.
{ 
In our model, a dynamically generated $\eta^\prime p$ bound state begins to appear for more attractive interaction with $\alpha \lesssim -1.2$ at $q_\text{max}=630$~MeV. 
The relation between the scattering length and the potential strength obtained in our calculation resembles that reported in Table~III in Ref.~\cite{Liu:2023uly}.
When we perform calculations with stronger attractive interactions, the value of the scattering length changes its sign and again begins to approach the experimental values of Eqs.~\eqref{eq:Exp1} and \eqref{eq:Exp2}. 

 }

\begin{table}[!tb]
\centering
\caption{The $\eta^\prime p$ scattering length $a_{\eta^\prime p}$ for the different values of the potential parameter $\alpha$. The positive (negative) sign of $\text{Re}\, a_{\eta^\prime p}$ corresponds to the attractive (repulsive) interaction.}
\begin{tabular}{c| c c }
\hline 
$\alpha$   &~~ $a_{\eta^\prime p}$ [fm] ~~ & ~~ $|a_{\eta^\prime p}|$ [fm]~~   \\
\hline
 $4.00$  &~~ $-0.406 + i~0.007$ ~~ & ~~ $0.406$~~   \\
 $0.50$  &~~ $-0.159 + i~0.010$ ~~ & ~~ $0.159$~~   \\
 $0.00$  &~~ $-0.002 + i~0.020 $ ~~ & ~~ $0.020$~~   \\
 $-0.25$  &~~ $0.141 + i~0.035 $ ~~ & ~~ $0.145$~~   \\
 $-0.50$  &~~ $0.390 + i~0.072 $ ~~ & ~~ $0.397$~~   \\
 $-0.55$  &~~ $0.464 + i~0.087 $ ~~ & ~~ $0.472$~~   \\
\hline
\end{tabular}
\label{tab:scat}
\end{table}

\begin{figure}[!tb]
\centering
\includegraphics[width=0.5\textwidth]{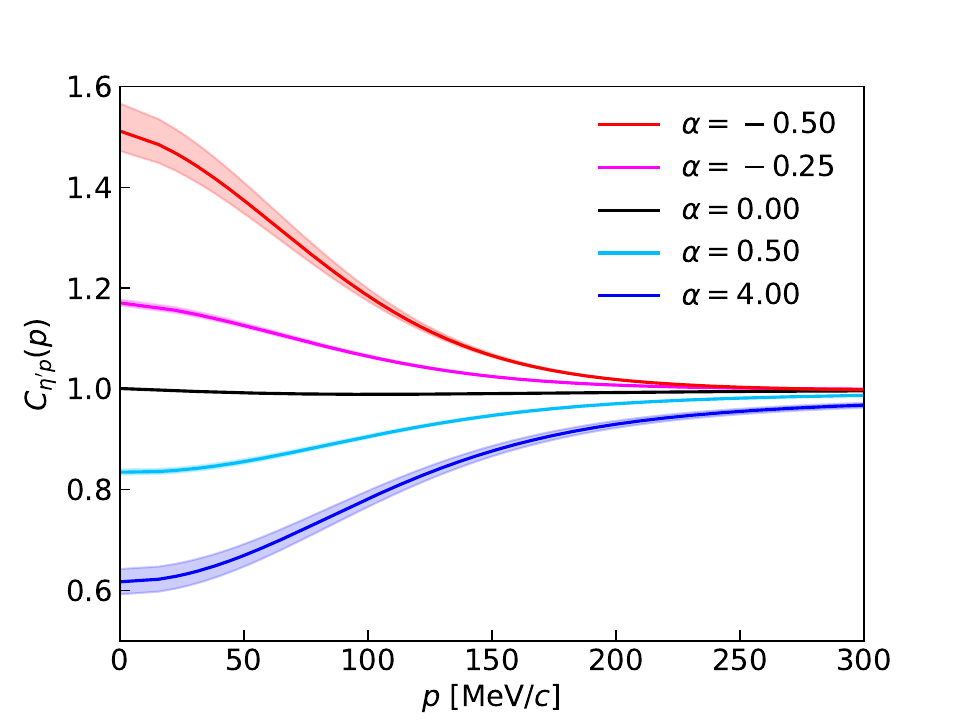}
\caption{The $\eta^\prime p$ correlation function for the different values of the potential parameter $\alpha$ in Eq.~\eqref{eq:Pot_alpha}. 
{The solid lines indicate the results calculated with $q_\text{max} = 630$~MeV, and the shaded bands show the results for $q_\text{max} = 637 \pm 72$~MeV.}
}
\label{fig:1}
\end{figure}

In Fig.~\ref{fig:1} we show the calculated $\eta^\prime p$ correlation function for different values of the parameter $\alpha$ in Eq.~\eqref{eq:Pot_alpha} corresponding to the different scattering lengths shown in Table~\ref{tab:scat}.
{
We also estimate the uncertainties of the correlation function by varying the cutoff parameter $q_{\text{max}}$ within the range $q_\text{max} = 637 \pm 72$~MeV evaluated in Ref.~\cite{Molina:2023jov}, and show the results as shaded bands in Fig.~\ref{fig:1}.
}
From the figure, we find that the $\eta^\prime p$ correlation function has largely different behaviors depending on the different
values of the potential parameter $\alpha$
and consequently different values of the scattering length $a_{\eta^\prime p}$.
We can see that the $\eta^\prime p$ correlation functions we obtained here have the typical feature of the attractive and repulsive interaction~\cite{Fabbietti:2020bfg,Liu:2023uly}
and have larger values than 1 for the attractive potential between the $\eta^\prime$ and the proton as shown in Fig.~\ref{fig:1}.
We also find that the $\eta^\prime p$ correlation function with the attractive interaction has a similar behavior to the $\eta p$ correlation function shown in Ref.~\cite{Molina:2023jov}.

\begin{figure}[!tb]
\centering
\includegraphics[width=0.5\textwidth]{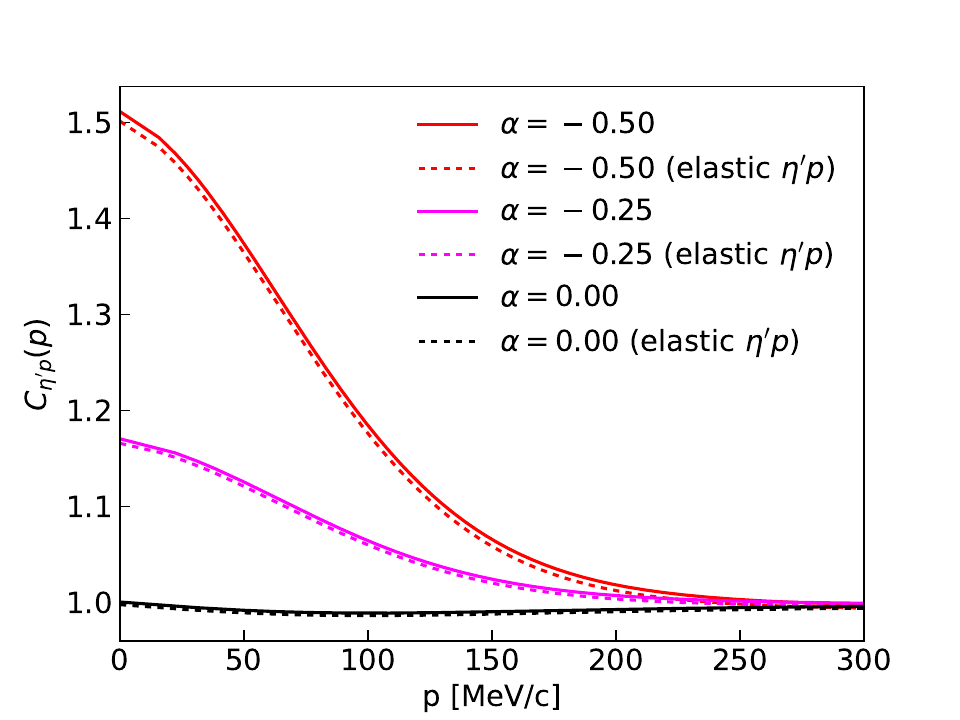}
\caption{ 
The $\eta^\prime p$ correlation functions for the elastic $\eta^\prime p$ channel $C^\text{el}_{\eta^\prime p}$ (dashed lines), and for the full coupled channels $C_{\eta^\prime p}$ (solid lines). 
The results indicated by the solid lines here are identical to those in Fig.~\ref{fig:1} for $\alpha=-0.50$, $-0.25$, $0.00$, respectively.
}
\label{fig:2}
\end{figure}

\begin{figure}[!tb]
\centering
\includegraphics[width=0.5\textwidth]{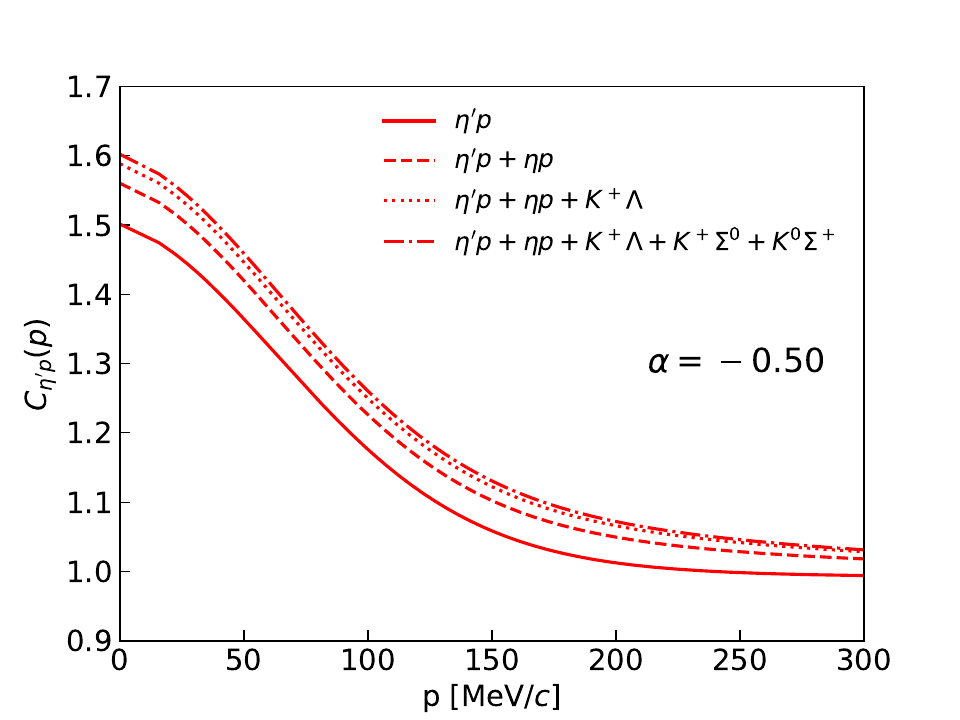}
\caption{ 
{
The $\eta^\prime p$ correlation functions for the different transitions to the $\eta^\prime p$ channel with the potential parameter $\alpha=-0.50$. 
The result labeled as $\eta^\prime p$ is identical to the elastic $\eta^\prime p$ channel $C^\text{el}_{\eta^\prime p}$ in Fig.~\ref{fig:2}. For each inelastic channel, the production weight of $\omega^{\text{prod}}_{i}=10$ is taken into account.
}
}
\label{fig:3}
\end{figure}

In Fig.~\ref{fig:2}, we also show the $\eta^\prime p$ correlation functions for the elastic channel $C^\text{el}_{\eta^\prime p}$ in Eq.~\eqref{eq:CF2} and for the fully coupled channels $C_{\eta^\prime p}$ in Eq.~\eqref{eq:CF} for the three different potential parameters $\alpha=-0.50, -0.25, 0.00$, which correspond to mostly attractive potentials.
{
Furthermore, we perform the calculations with large production weights $\omega^{\text{prod}}_{i}$ ($i = \eta p, K^+ \Lambda, K^+ \Sigma^0, K^0 \Sigma^+$) for inelastic channels, ranging from 1 to 10, following the analysis reported in Refs.~\cite{Encarnacion:2024jge,Feijoo:2024bvn}.
In Fig.~\ref{fig:3}, we show the contributions of the different transitions to the $\eta^\prime p$ with the potential parameter $\alpha=-0.50$. 
For each inelastic channel, we set the production weight to $\omega^{\text{prod}}_{i}=10$ as an upper limit value, which corresponds to the largest estimate in Ref.~\cite{Encarnacion:2024jge}.
Even in this extreme case, the effect of the inelastic channels modifies the correlation function by less than about 10\%, and the elastic contribution $C^\text{el}_{\eta^\prime p}$ remains dominant.
} 
We find that the $\eta^\prime p$ correlation function is mostly determined by the elastic channel and is not affected much by other inelastic channels.
This is an important finding because we do not know exactly the weights $\omega^{\text{prod}}_{i}$ of the inelastic channels which are assumed to be 1 here as mentioned after Eq.~\eqref{eq:CF}.
Therefore, we can conclude that 
the measurement of the $\eta^\prime p$ correlation function provides a valuable information on the $\eta^\prime p$ interaction. \\

\section{Summary and Conclusions} \label{sec:discussion}
We have studied the correlation function of the $\eta^\prime N$ system to study the $\eta^\prime N$ interaction. 
The $\eta^\prime N$ interaction is a crucial subject of theoretical and experimental study, and it is necessary to settle the issue about the possibility of having $\eta^\prime$--nucleus bound states, which has attracted much interest in the nuclear physics community. 
The measurement of the $\eta^\prime p$ correlation function could come to the rescue, and in the present work, we have faced this problem from the theoretical point of view. 

We have found that the $\eta^\prime p$ correlation function is very sensitive to the $\eta^\prime N$ interaction and the value of the $\eta^\prime p$ scattering length, which is very poorly known at present.  
One problem in principle would be the unknown effect of the coupled channels, which is reflected in weights carried by the inelastic channels in the calculation of the correlation function. 
However, in the present work, we found that the effect of these inelastic channels in the $\eta^\prime p$ correlation function is extremely small. 
As a result, the measure of the $\eta^\prime p$ correlation function alone can provide very valuable information on the $\eta^\prime p$ scattering length, and indirectly the strength of the $\eta^\prime N$ interaction. 
The large sensitivity of the correlation function to the still unknown $\eta^\prime p$ scattering length, makes the correlation function ideal to learn about this scattering length.
We hope that the findings of the present work stimulate the experimental search for the $\eta^\prime p$ correlation function, which would provide a much needed information to make progress on the issue of the $\eta^\prime N$ interaction and the possible $\eta^\prime $ bound states in nuclei indicated by very recent experimental data~\cite{Sekiya:2025hwz}.

\section*{Acknowledgments}
We would like to thank E. Oset, D. Jido, S. Hirenzaki, Y. Kamiya for valuable discussions.
The work was partly supported by JSPS KAKENHI Grant Numbers 23K03417  and JP24K07020.

\bibliography{ref_etaprime}

\end{document}